\RequirePackage{lineno} 
\documentclass[prd, reprint, preprintnumbers, twocolumn, eqsecnum,floatfix,letterpaper,superscriptaddress,nofootinbib]{revtex4-1}
\usepackage{graphicx,url,amssymb,amsmath,rotating,color,units,wasysym,epsfig,multirow,epstopdf,subcaption}
\usepackage[colorlinks,urlcolor=blue,citecolor=blue,linkcolor=blue]{hyperref}
\usepackage{comment}
\usepackage{lineno}
\graphicspath{{./}{figures/}}

\begin{document}

\title{A machine-learning pipeline for real-time detection of gravitational waves from compact binary coalescences}
\author{Ethan Marx}
\affiliation{Department of Physics, MIT, Cambridge, MA 02139, USA}
\affiliation{LIGO Laboratory, 185 Albany St, MIT, Cambridge, MA 02139, USA}
\author{William Benoit}
\affiliation{School of Physics and Astronomy, University of Minnesota, Minneapolis, MN 55455, USA}
\author{Alec Gunny}
\affiliation{Department of Physics, MIT, Cambridge, MA 02139, USA}
\affiliation{LIGO Laboratory, 185 Albany St, MIT, Cambridge, MA 02139, USA}
\author{Rafia Omer}
\affiliation{School of Physics and Astronomy, University of Minnesota, Minneapolis, MN 55455, USA}
\author{Deep Chatterjee}
\affiliation{Department of Physics, MIT, Cambridge, MA 02139, USA}
\affiliation{LIGO Laboratory, 185 Albany St, MIT, Cambridge, MA 02139, USA}
\author{Ricco C. Venterea}
\affiliation{School of Physics and Astronomy, University of Minnesota, Minneapolis, MN 55455, USA}
\affiliation{Department of Astronomy, Cornell University, Ithaca, NY, 14853, USA}
\author{Lauren Wills}
\affiliation{School of Physics and Astronomy, University of Minnesota, Minneapolis, MN 55455, USA}
\author{Muhammed Saleem}
\affiliation{Physics Department, University of Texas at Austin, Austin TX 78712 USA}
\affiliation{School of Physics and Astronomy, University of Minnesota, Minneapolis, MN 55455, USA}

\author{Eric Moreno}
\affiliation{Department of Physics, MIT, Cambridge, MA 02139, USA}
\affiliation{LIGO Laboratory, 185 Albany St, MIT, Cambridge, MA 02139, USA}
\author{Ryan Raikman}
\affiliation{Department of Physics, MIT, Cambridge, MA 02139, USA}
\affiliation{Department of Physics, Carnegie Mellon University, Pittsburgh, PA, 15213}
\author{Ekaterina Govorkova}
\affiliation{Department of Physics, MIT, Cambridge, MA 02139, USA}
\affiliation{LIGO Laboratory, 185 Albany St, MIT, Cambridge, MA 02139, USA}
\author{Malina Desai}
\affiliation{Department of Physics, MIT, Cambridge, MA 02139, USA}
\affiliation{LIGO Laboratory, 185 Albany St, MIT, Cambridge, MA 02139, USA}

\author{Jeffrey Krupa}
\affiliation{Department of Physics, MIT, Cambridge, MA 02139, USA}

\author{Dylan Rankin}
\affiliation{Department of Physics and Astronomy, University of Pennsylvania, Philadelphia, PA, 19104, USA}
\author{Michael W. Coughlin}
\affiliation{School of Physics and Astronomy, University of Minnesota, Minneapolis, MN 55455, USA}
\author{Philip Harris}
\affiliation{Department of Physics, MIT, Cambridge, MA 02139, USA}
\author{Erik Katsavounidis}
\affiliation{Department of Physics, MIT, Cambridge, MA 02139, USA}
\affiliation{LIGO Laboratory, 185 Albany St, MIT, Cambridge, MA 02139, USA}

\date{\today}


\begin{abstract}
The promise of multi-messenger astronomy relies on the rapid detection of gravitational waves at very low latencies ($\mathcal{O}$(1\,s)) in order to maximize the amount of time available for follow-up observations.
In recent years, neural-networks have demonstrated robust non-linear modeling capabilities and millisecond-scale inference at a comparatively small computational footprint, making them an attractive family of algorithms in this context.
However, integration of these algorithms into the gravitational-wave astrophysics research ecosystem has proven non-trivial.
Here, we present a machine learning-based pipeline for the detection of gravitational waves from compact binary coalescences (CBCs) designed to run in low-latency. 
We demonstrate this pipeline to have a fraction of the latency of traditional matched filtering search pipelines while achieving state-of-the-art sensitivity to higher-mass stellar binary black holes.
\end{abstract}
 \maketitle

\section{Introduction} \label{sec:intro}

Gravitational-wave astronomy has developed rapidly since the first direct detection of gravitational waves from a binary black hole merger in 2015\cite{AbEA2016a}, with new detections now a common occurrence\cite{Abbott_2023}.
With the fourth observing run (O4) of the LIGO-Virgo-KAGRA (LVK) collaboration \cite{Aasi_2015, Acernese_2015, kagra} already underway, and with future ground and space based detectors planned for various points in the next decade\cite{PhysRevD.91.082001, Punturo_2010, LISA2000}, ever more frequent discoveries of gravitational waves will enable follow-up observation of events across other cosmic messengers such as electromagnetic radiation and astrophysical neutrinos\cite{Tanvir_2013,Ascenzi_2019,Jin_2020,Rastinejad_2022,Albert_2023,Abbasi_2023}. 
The insights we gain in this era of multi-messenger astrophysics will directly correlate with the volume and diversity of data we are able to collect.
Already, the first multi-messenger event, which included the gravitational wave event GW170817~\citep{AbEA2017b}, the gamma ray burst GRB170817A~\citep{GoVe2017,SaFe2017,AbEA2017e,Savchenko_2017}, and the kilonova AT2017gfo~\citep{CoFo2017,SmCh2017,AbEA2017f} has led to a gold mine of new science.

While machine learning (ML) is ubiquitous in some areas of physics\cite{harris2022physics}, it has only recently approached a stage of maturity in the gravitational-wave community.
To date, there have been a number of machine learning models proposed for the detection of compact binary coalescences (CBCs); e.g.,\cite{PhysRevD.103.102003, verma2022can, KRASTEV2020135330, PhysRevD.97.044039, nousi2022deep}; but there are none currently running in O4\cite{online_pipelines} (though, ML-based unmodeled gravitational-wave searches have seen production usage\cite{mly}).
This is both a product of well-known infrastructure hurdles separating the development and deployment of machine learning models\cite{deployment_challenges}, as well as a lack of standardized, astrophysically meaningful probes of the sensitivity of these models in the face of non-stationary and transient background noise.

The most well-modeled and frequently observed gravitational-wave events to date are the mergers of binary black hole (BBH) systems\cite{AbEA2018b,theligoscientificcollaboration2022gwtc21,Abbott_2023}
Their comparatively high number of confirmed detections has given us reasonable models of their population statistics, allowing for astrophysically meaningful measures of search sensitivity. BBH mergers also benefit from a highly localized-in-time signal-to-noise ratio (SNR)\footnote{{In the remainder of this paper, SNR refers to the total SNR of a Hanford and Livingston detector network}} profile relative to binary neutron star (BNS) mergers, which are in the sensitive band of the detectors much longer. Studying the ability of neural-networks to detect BBH mergers, and in particular what real time use in the LVK detectors looks like in this context, represents an important first step towards developing a more thorough understanding of how, and whether, these algorithms can be applied to more challenging signals such as BNSs, and what tools and infrastructure would be required to do so.

Here, we present \texttt{Aframe}, a flexible pipeline for detection of BBH mergers using deep learning.
The implementation presented here uses a 1D convolutional neural-network. Convolutional neural-networks have previously been shown to have potential for gravitational wave detection\cite{magicbullet}, and we use this architecture, along with aggressive data augmentation techniques, to achieve a sensitivity competitive with matched filtering CBC search pipelines while requiring a significantly lower latency.
More broadly, \texttt{Aframe} encompasses a suite of tools for quickly implementing, testing, and deploying new ideas at scale in order to more confidently realize the potential of machine learning in service to gravitational wave astronomy.

The structure of this paper is as follows. Sec.~\ref{sec:aframe-overview} gives a high-level overview of the \texttt{Aframe} algorithm. In Sec.~\ref{sec:sv-calc}, we describe the metric we use to measure  performance. Sec.~\ref{sec:data} describes the datasets used to train and evaluate our network, and the means by which training and evaluation is performed is given in Secs.~\ref{sec:training} and \ref{sec:inference}, respectively. We discuss the longevity of our model in Sec.~\ref{sec:longevity} and the latency and computational requirements in Sec.~\ref{sec:compute}. Finally, we compare our performance to existing pipelines in Sec.~\ref{sec:comparison} and examine subset of GWTC-3 catalog events in Sec.~\ref{sec:catalog-events}

\section{The \texttt{Aframe} algorithm} \label{sec:aframe-overview}

Our neural-network architecture modifies a standard ResNet34\cite{resnet}, which maps fixed length time-series of gravitational wave strain from two interferometers (here, the Hanford and Livingston LIGO interferometers) to a scalar detection statistic indicating whether a signal is present in the input. The detection statistic is analogous to the SNR output of matched filtering searches; the larger the value, the more likely the neural-network believes there is a signal consistent with the training distribution present in the data. Critically, we replace 2D with 1D convolutions to accommodate time-series input. In addition, we replace standard Batch Normalization layers (BN)\cite{ioffe2015batch}, with Group Normalization (GN) layers\cite{Wu_2018_ECCV}. While BN layers fit parameters to statistics calculated along the batch dimension, GN layers are fit to statistics calculated from groups of channels. This choice was motivated by differences in the statistical properties of batches during training and inference. During training, there are significantly more signals in each batch than during inference, where most of the batch consists of noise. Thus, during training, BN layers will learn spurious statistical properties that are not present at inference time. GN layers mitigate this problem by learning statistical properties of individual channels. We found that using GN layers improves the agreement between validation and test time metrics, as well as overall testing performance. Good agreement between validation and test metrics is essential for ensuring the best neural-network is being selected for deployment. The neural-network is trained by minimizing a binary cross entropy loss function with an Adam\cite{AdamOptimizer} optimizer. We use a one cycle learning rate scheduler with cosine annealing\cite{onecycleLR}. 

Analyzing data with \texttt{Aframe} involves loading and preprocessing timeseries data, breaking it up into short time segments, then passing these segments through the neural-network. The throughput associated with each of these steps can vary drastically, as can the hardware and software necessary to accelerate them. In order to optimize the total throughput of this system, we adopt an inference-as-a-service (IaaS) computing model in which neural-network inference is handled by a dedicated service, to which client applications can send inference requests remotely. Each step in our pipeline is then implemented and scaled independently to most efficiently leverage a fixed pool of heterogeneous computing resources. This model has been shown to be effective in optimizing ML inference in GW astronomy\cite{iaas}, provided that “snapshotting”\cite{hermes} is used to cache overlapping input data on the server side to avoid redundant data transfer. We adopt this paradigm using an off-the-shelf IaaS implementation, Triton Inference Server\cite{NVIDIA_Corporation_Triton_Inference_Server}, and use the ML inference framework TensorRT to accelerate the neural-network inference step. The ability to scale and distribute a workload is an important part of any search pipeline, and the authors are aware of only one other ML-based CBC detection algorithm that has focused on scalability to arbitrary resources\cite{huerta2021}. In the sections below, we compare both our sensitivity and our throughput to this work.

\begin{figure}[h]
\centering
\includegraphics[width=\columnwidth]{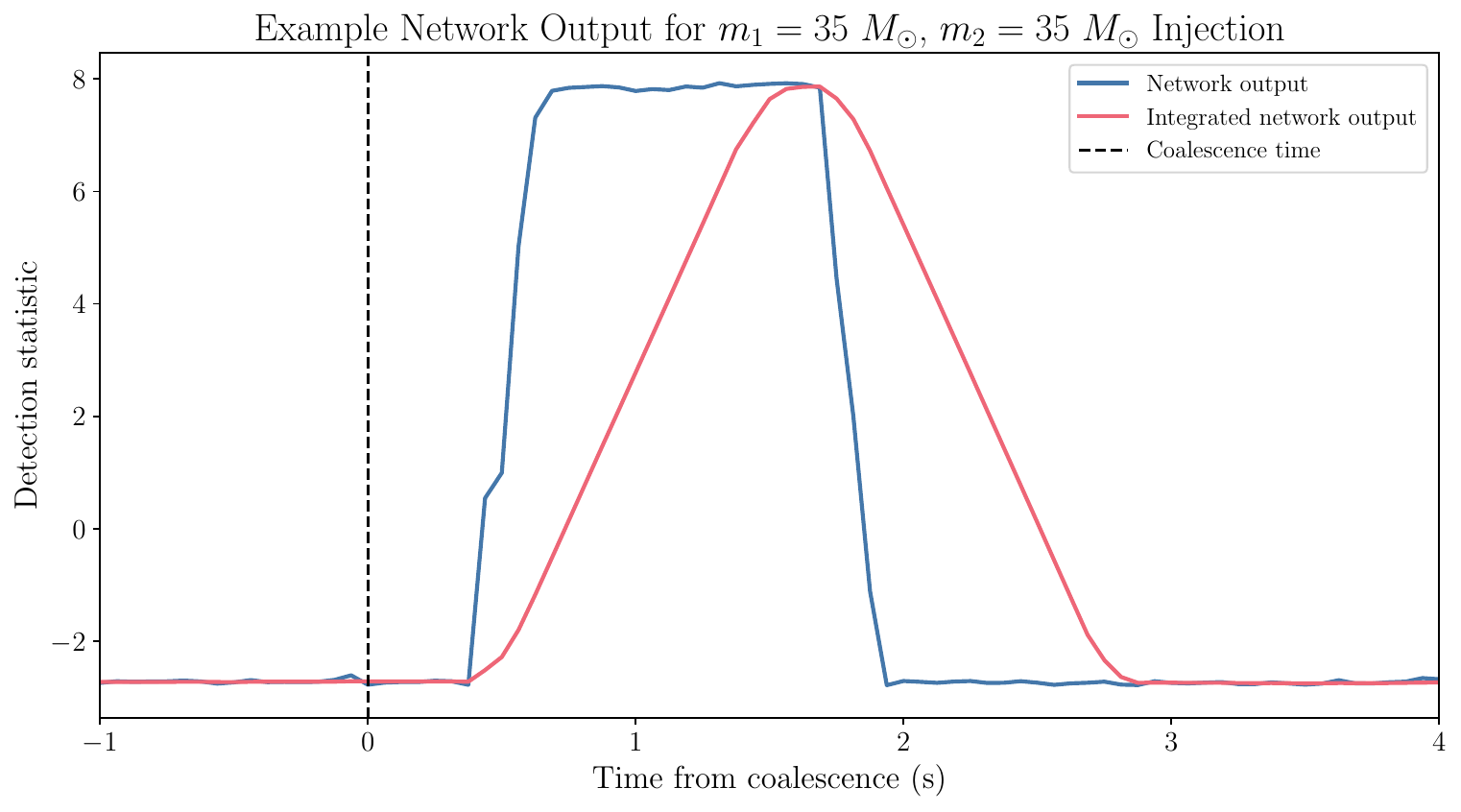}
\caption{Example neural-network prediction and integrated neural-network prediction for a $m_1 = 35 M_\odot, m_2 = 35 M_\odot$ signal injection. The coalescence time is plotted as the vertical dashed black line. The 0.75\,s gap between coalescence time and the full neural-network activation is due to the whitening filter settle-in time and the placement of injections during training; see Sec.~\ref{sec:noise-sampling}.}\label{fig:example-output}
\end{figure}

Inference is performed at a rate of 4\,Hz (not to be confused with the neural-network throughput, see the discussion of computational requirements below). In other words, we pass windows of data to our neural-network for inference such that each window is shifted by 0.25\,s. This inference sampling rate reduces the overall compute load without sacrificing search sensitivity (see Sec.~\ref{sec:inference}).
These neural-network predictions are then integrated over time using a 1\,s top hat filter (see Fig.~\ref{fig:example-output}). Because the neural-network is trained to encode time translation invariance (see Sec.~\ref{sec:signal-injections}), we expect to see a consistently high neural-network responses when analyzing astrophysical signals.
Thus, integration provides a mechanism to promote consistently high outputs while rejecting short transients that may correspond to non-astrophysical sources.
Finally, the integrated time-series of neural-network predictions is clustered to avoid yielding multiple triggers for the same event. The maximum integrated value over an 8\,s window is taken as the detection statistic  corresponding to a candidate event.
The coalescence time can then be estimated by accounting for the integration window length, whitening filter settle-in, and the placement of the coalescence time relative to the right edge of the kernel during training (see \ref{fig:example-output}).
The mean absolute time difference between the estimated coalescence time and the injected time for recovered signals in the testing set was 0.123\,s with a standard deviation of 0.125\,s, which is in line with the inference sampling rate of 4\,Hz.

\section{Sensitive Volume} \label{sec:sv-calc}
A key metric in understanding a search algorithm's performance is the \textit{sensitive volume}, which is a measure of the region of space in which a pipeline is expected to detect merging binaries.
The sensitive volume as a function of the FAR is defined by
\begin{equation}
    V(\mathcal{F}) = \int \mathrm{d}\mathbf{x}\ \mathrm{d} \theta\ \epsilon(\mathcal{F}; \mathbf{x}, \theta) \phi(\mathbf{x}, \theta)
\end{equation}
where $\phi$ is the distribution of events over spatial coordinates $\mathbf{x}$ and binary system parameters $\theta$, and $\epsilon$ is the detection efficiency of the pipeline at a false alarm rate $\mathcal{F}$\cite{PhysRevD.102.063015}.
Generally, this quantity is estimated using Monte-Carlo integration by drawing waveforms from a population model, injecting them into detector strain data, and counting how many produce triggers below a given false alarm rate threshold.
If the samples are drawn from within the redshifted volume\cite{Chen_2021} $V_0$, with
\begin{equation}
    V_0 = \int_{z_{\text{min}}}^{z_{\text{max}}} \mathrm{d}z \frac{\mathrm{d}V_c}{\mathrm{d}z} \frac{1}{1+z} 
\end{equation}
where $\mathrm{d}V_c/\mathrm{d}z$ is the differential comoving volume, then the sensitive volume is approximately
\begin{equation}
    V(\mathcal{F}) \approx V_0 \frac{N(\mathcal{F})}{N_{\text{draw}}}
\end{equation}
where $N(\mathcal{F})$ is the number of signals detected at a FAR less than $\mathcal{F}$ and $N_{\text{draw}}$ is the number of injected events.

It is often desired to quantify the sensitivity of an algorithm to different populations. For example, an algorithm's sensitivity may vary with different source masses. Through the technique of importance sampling, it is possible to use one injection set from a broad population to calculate the sensitive volume for several populations. Each injection is weighted by the ratio of the probability of having been drawn from the injected distribution to that of the population distribution of interest\cite{Tiwari_2018}:
\begin{equation}
    V_{\text{pop}}(\mathcal{F}) \approx \frac{V_0}{N_{\text{inj}}} \sum_{i=1}^{N(\mathcal{F})} \frac{p_{\text{pop}}(\theta_i)}{p_{\text{inj}}(\theta_i)}
\end{equation}
The Monte-Carlo uncertainty on this estimation is\cite{Farr_2019}

\begin{equation}
    (\delta V_{\text{pop}})^2 = \frac{V_0^2}{N_{\text{inj}}^2} \sum_{i=1}^{N(\mathcal{F})} \left(\frac{p_{\text{pop}}(\theta_i)}{p_{\text{inj}}(\theta_i)} \right)^2 - \frac{V_{\text{pop}}^2}{N_{\text{inj}}}
\end{equation}
The SNR-based rejection performed during the generation of test set waveforms is done to improve this uncertainty.
Waveforms that are sampled but have an SNR less than 4 are not injected; however, they are still counted towards $N_{\text{draw}}$.
The cut is placed such that any waveforms below the SNR cutoff are not expected to be recovered at any reasonable FAR, and so would not contribute to the sensitive volume: whether injected or not, their weight would be zero.
This procedure allows us to effectively draw many times more samples than are actually injected, greatly reducing the uncertainty on the sensitive volume.
For this analysis, we re-weight to the same population distributions used in the sensitive volume analysis conducted in GWTC-3\cite{Abbott_2023}, log-normal distributions about central masses of interest with widths of 0.1. In addition, we enforce time difference of no more than 0.25\,s between the recovered and injected coalescence times. This time difference corresponds to the resolution available at an inference sampling rate of 4\,Hz. This time resolution can be reduced by increasing the inference sampling rate.

\section{Data} \label{sec:data}
\subsection{Strain}
We train and validate our neural-network using open data from the Gravitational Wave Open Science Center (GWOSC)\cite{Abbott_2023_open_data} between times 2019-04-29T13:29:25 and 2019-05-09T13:29:25, corresponding to a ten calendar day period at the beginning of the O3 observing run. The strain data is resampled to 2048\,Hz for better computational efficiency. For each interferometer, we query the openly available science mode flag to remove segments with poor data quality. We then select segments for which the science mode flag is active for both the Hanford and Livingston LIGO interferometers. This amounts to approximately 4.7 days of coincident livetime. We reserve the segments that total a minimum of 15,000 seconds at the end of this period for validating the neural-network throughout the training process.

For evaluating the performance reported in Fig.~\ref{fig:sensitivity}, we select data satisfying the above criteria between times 2019-05-09T13:29:25 and 2019-06-08T13:29:25, corresponding to a 30 day period immediately after the training period. This amounts to approximately 18 days of coincident livetime. During evaluation, timeslides (see Sec.~\ref{sec:comparison}) of this data are created such that the total desired background livetime is achieved. We emphasize that no data used for evaluating the performance of the neural-network was used during training or validation. In addition, we train the neural-network only with data from before the testing period. This mimics the data availability scenario for real-time application. 

\subsection{Waveforms} 
We use \texttt{bilby}\cite{Ashton_2019} to simulate 100,000 eight second long BBH waveforms at 2048\,Hz with the IMRPhenomPv2 approximant\cite{PhysRevLett.113.151101}. Out of these, 75,000 waveforms are used to train the neural-network, and the remaining 25,000 are reserved for validation. To simulate a waveform, a probability distribution is specified on each of the parameters that define a compact binary merger, and random samples are drawn from each.
The distribution set used in this work is based on one used for GWTC-3\cite{PhysRevX.13.011048} during O3 to assess the sensitivity of CBC search pipelines, and is described in Table~\ref{table:priors}. 
The sampled parameters are used to compute the time-domain strain for each polarization, $h_+$ and $h_\times$. The sampled component mass values are defined in the source frame, so conversion to detector frame quantities is performed before generation.
The interferometer responses of the intrinsic polarizations are calculated during the training process to allow for real-time data augmentations, as described below in Sec.~\ref{sec:training}.

The same distribution (Table~\ref{table:priors}) is used to simulate signals for the testing dataset. Enough waveforms are generated to fill the background timeslides with the waveform coalescence points spaced 24\,s apart. As the signals are only 8\,s long, they do not overlap. Because waveforms are distributed uniformly in co-moving volume, the majority of waveforms are placed at high redshifts, and thus the population is dominated by low SNR signals. So, during the signal generation process, we perform rejection sampling and keep only signals that have an SNR greater than 4. This ensures that we can accurately evaluate sensitivity to an astrophysically motivated population without wasting computation on signals we do not expect to detect\cite{ligo_scientific_collaboration_and_virgo_2023_7890437}. Rejection sampling reduces the uncertainty of a sensitive volume estimate for a fixed amount of analyzed injections (see Sec.~\ref{sec:sv-calc}). In total, we generate $\sim 45,000,000$ waveforms. Of these, $\sim3\%$ percent are used for testing and $\sim97\%$ are rejected. 

We apply several data augmentation techniques during the training process with the goal of providing robust, high entropy data that encodes physics-based knowledge for discriminating signals from noise. 
Below, we will describe how a training batch is composed, as well as the hyper-parameters that control the composition of the batches. 

\textbf{Noise sampling.} \label{sec:noise-sampling}
Sampled at 2048\,Hz, the entire training dataset is unable to fit onto a single 16\,GB V100 GPU at once. 
Thus, efficient out-of-memory data-loading is required to fully utilize the extent of our strain dataset. To do this, we sample strain windows directly from disk during the training procedure. The length of each noise window sampled from disk is 10.5\,s. The first 8\,s is used to estimate the power spectral density (PSD) used for whitening. The remaining 2.5\,s of the window is whitened in the frequency-domain, and transformed back to time-domain. Due to whitening filter settle-in, 0.5\,s of data is corrupted on both ends of the window and removed. Thus, only 1.5\,s of data is actually analyzed by the neural-network. The PSD estimation, filter construction, and whitening are all done with \texttt{PyTorch}\cite{10.5555/3454287.3455008} modules to enable GPU-accelerated computation\cite{nousi2022deep}. We use a training batch size of 384, which was chosen such that we fully utilize the GPU memory available. Our out-of-memory data-loading is sufficiently fast to support these batch sizes without bottle-necking the pre-processing or neural-network modules.

Noise instances are sampled independently in time from each interferometer.
Thus, a noise instance from one interferometer can be paired with many different instances from the other interferometer.
This combinatorially increases the amount of unique two-detector noise instances available for optimizing the network. 
Next, each noise instance has probability $p_{\text{invert}}$ to be inverted ($h(t) \rightarrow -h(t)$) and, independently, probability $p_{\text{reverse}}$ to be reversed ($h(t) \rightarrow h(-t)$)\cite{Bini_2023}.
Again, the inversion and reversal augmentations increase the amount of unique noise instances in our training data.
For transient noise, these augmentations increase the variety of morphologies provided during training, allowing for better generalization to unseen testing data. 
We fix $p_{\text{invert}}$ and $p_{\text{reverse}}$ to 0.5.

\textbf{Signal Injection.} \label{sec:signal-injections}
Once a batch of noise instances is generated, simulated BBH signals are added into each 2.5\,s unwhitened window with probability $p_{\text{signal}} = 0.277$ and labeled as signals; this signal probability is one of six hyperparameters that we search over (see Table~\ref{table:hp-priors} and the discussion of hyperparameters below). 
The procedure for injecting signals is as follows: first, intrinsic polarization time-series are randomly sampled from the training waveform bank. 
Next, random extrinsic parameters (right ascension, declination, polarization angle, and SNR) are sampled. The first three of these are sampled from the priors described in Table~\ref{table:priors}; We will discuss the method of SNR sampling in the following paragraph.
Intrinsic polarization time-series are then projected onto the interferometers and re-scaled to the sampled SNR. 
Randomly sampling extrinsic parameters at training time allows each intrinsic time-series to be injected from a variety of sky localizations and distances throughout the training procedure. We found that standard CPU implementations of projecting intrinsic polarizations onto interferometers created bottlenecks that severely limited utilization of GPU resources. We eliminated this bottleneck by developing a \texttt{PyTorch}\cite{10.5555/3454287.3455008} implementation so that projection can be accelerated using GPUs by a factor of $\sim$\,200.
Finally, the interferometer responses are added into the noise instances.
The coalescence time of the merger is randomly placed so that it falls at least 0.25\,s from either edge of the 1.5\,s whitened noise instance.
We enforce this padding because we found that having the coalescence point too close to the left edge of the window makes it more difficult for the neural-network to learn, since much of the signal SNR would lie outside the window.
The random placement of the coalescence time encodes time translational invariance so that the neural-network can identify signals with the coalescence time at different locations throughout the window.

\begin{table}
\centering
\footnotesize
\begin{tabular*}{\columnwidth}{ p{4cm} p{1.5cm} p{1.75cm} p{1.5cm} }
\hline
\hline
Parameter & Prior & Limits & Units\\
\hline
Mass of primary & $m_1^{-2.35}$ & $(5, 100)$ & $M_{\odot}$\\
Mass of secondary & $m_2$ & $(5, m_1)$ & $M_{\odot}$\\
Redshift & Comoving & $(0, 2)$ & -\\
Polarization angle & Uniform & $(0, \pi)$ & rad.\\
Dimensionless spin magnitude & Uniform & $(0, 0.998)$ & -\\
Spin tilt & Sine & $(0, \pi)$ & rad.\\
Relative spin azimuthal angle & Uniform & $(0, 2\pi)$ & rad.\\
Spin phase angle & Uniform & $(0, 2\pi)$ & rad.\\
Orbital phase & Uniform & $(0, 2\pi)$ & rad.\\
Right ascension & $(0, 2\pi)$ & rad.\\
Declination &Cosine & $(-\pi/2, \pi/2)$ & rad.\\
Inclination angle & Sine & $(0, \pi)$ & rad.\\
\hline
\hline
\end{tabular*}
\caption{Priors on parameters used to generate waveforms for both the training and testing sets. The prior is derived from that used in GWTC-3 to assess search pipelines. The component mass distributions are defined in the source frame. ``Comoving" refers to uniform in comoving volume.}
\label{table:priors}
\end{table}

\section{Training}
\label{sec:training}

\textbf{Curriculum Learning.}
Curriculum learning is a technique for training machine learning models in which initially, easy to learn samples are provided as training data, and progressively harder samples are introduced over time.
One way to apply this in the context of GW detection is to initially provide high SNR signals and gradually introduce lower SNR signals\cite{nousi2022deep}.
This allows the neural-network to quickly arrive at a minima of its parameter space before trying to optimize for the more realistic task.
We begin with an SNR distribution that follows a power law, $p(\text{SNR}) \sim (\text{SNR})^{-3}$, with a minimum of SNR$_{\text{min}} = 12$ and a maximum of SNR$_{\text{max}} = 100$.
The form of this distribution was chosen to roughly match the SNR distribution of of our astrophysically motivated prior. 
Each time a new training batch is constructed, the minimum SNR bound of the distribution is decreased until we reach the ultimate lower bound of 4.
This decrease happens uniformly over 989 batches, a value that was reached through a hyperparameter search.

\textbf{Glitch Mitigation.}
Non-Gaussian noise transients, known as ``glitches,'' can often mimic BBH signals and lead to high-significance false alarms. We implement two types of augmentations we call waveform \emph{muting} and \emph{swapping} to mitigate the impact of transient glitches. These augmentations respectively encode the concepts of coincidence and coherence that true astrophysical signals are expected to exhibit. The values of the parameters controlling these augmentations were determined by hyperparameter search; see below for more details.

\emph{Muting}: For a fraction $p_{\text{mute}} = 0.055$ of the training batch, we inject a BBH signal into only one of the interferometers and label these samples as noise. This teaches the neural-network that it is not enough for a BBH-like signal to be present in just one interferometer: coincidence between interferometers is a requirement for true astrophysical signals. 

\emph{Swapping}: For an independent fraction of the training batch, $p_{\text{swap}} = 0.014$, we swap one of the interferometer responses with an interferometer response from different signal, and label these samples as noise. Thus, these windows will contain BBH waveforms with different intrinsic parameters in each interferometer. This motivates the neural-network to learn the concept of coherence: the time-frequency evolution of the signal must be identical in both interferometers.



\begin{table*}
\centering
\small
\begin{tabular*}{\textwidth}{ p{2cm} p{8.5cm} p{2.5cm} p{2.5cm} p{2.5cm} }
\hline
\hline
Parameter & Description & Prior & Limits & Best Value \\
\hline
$lr_{\text{max}}$ & Maximum learning rate & Log Uniform  &  $(10^{-4.5}, 10^{-2})$  & $5.8\times10^{-4}$\\
$N_{\text{ramp}}$ & Number of epochs over which learning rate increases & Uniform & $(2, 50)$ & $23$ \\
$p_{\text{signal}}$ & Probability of batch element containing a signal & Uniform & $(0.2, 0.6)$ & $0.277$\\
$p_{\text{swap}}$ & Probability of swap augmentation & Uniform & $(0, 0.15)$ & $0.014$\\
$p_{\text{mute}}$ & Probability of mute augmentation & Uniform & $(0, 0.3)$ & $0.055$ \\
SNR steps & Number of batches over which SNR scheduler decays & Uniform & $(1, 2500)$ & $989$\\
\hline
\hline
\end{tabular*}
\caption{Priors and descriptions of hyperparameters searched over. The best value corresponds to the neural-network from the hyperparameter search that produced the highest validation score across all epochs. A neural-network trained with these hyperparameters was used to evaluate results reported in Fig.~\ref{fig:sensitivity}. Details on hyperparameters can be found in Sec.~\ref{sec:hp-search}}
\label{table:hp-priors}
\end{table*}

\textbf{Validation.}
We construct our validation procedure with the goal of establishing a strong correlation between validation and test metrics. This allows us to confidently pick the best performing neural-network during a hyperparameter search, as well as during individual training runs. To accomplish this, our validation procedure is designed to mimic the testing procedure as closely as possible. We reserve 15,000 seconds of strain data from immediately after the training period and 25,000 waveforms exclusively for neural-network validation during training. This data is not used at all for training the neural-network. This temporal choice of training and validation split mimics the real-time production setting, where a deployed neural-network is only trained on past data. 

To construct our validation set, we first create timeslides of the background data until at least 16 hours of livetime is accumulated. Similarly to training, this data is batched into 10.5\,s windows, with the first 8\,s used for whitening the final 2.5\,s of each window. As with the training data, 0.5\,s of data is cropped from each edge of the window after whitening.
Next, we create a dataset of injections by adding waveforms from the validation waveform dataset into the background windows. We set a minimum detector-network SNR threshold of 4 for validation signals. Signals that are quieter are re-scaled to the SNR 4 threshold. The SNR is computed with respect to the PSD calculated from the first 8\,s of the window. This rescaling procedure mimics the SNR-based rejection sampling performed for the testing dataset. We create 5 unique injection sets that have the coalescence point of each waveform at 0.25, 0.5, 0.75, 1.0, and 1.25\,s within each whitened window. This ensures the validation metric covers a wider variety of scenarios. 

The neural-network outputs a prediction for each window in the background and injection datasets. We use these predictions to calculate the area under the ROC curve (AUROC) up to a false positive rate (FPR) of $10^{-3}$, which is the final validation metric. We make this cut on the AUROC so that we are optimizing performance in the regime of low FARs. After the neural-network training has converged, the weights corresponding to the epoch with the highest validation score are used for testing.

\textbf{Hyperparameter Search.} \label{sec:hp-search}
The hyperparameters of our algorithm are optimized via a random search\cite{random_search}. It is infeasible to search over all possible hyperparameters, so we selected those that we a-priori expect to have the greatest impact on the neural-network optimization process.
These were the neural-network's maximum learning rate ($lr_{\text{max}}$), the number of epochs over which the learning rate ``ramps up" ($N_{\text{ramp}}$) to $lr_{\text{max}}$ , $p_{\text{signal}}$, $p_{\text{mute}}$, $p_{\text{swap}}$, and the number of steps over which SNR curriculum learning was performed. The priors on each of these parameters can be found in Table~\ref{table:hp-priors}.
30 combinations of these parameters were randomly sampled and used to train a neural-network.
Of these, the neural-network that reported the highest validation score was selected as the neural-network used for testing. The hyperparameters used to train this neural-network are reported in Table~\ref{table:hp-priors}.

\section{Inference} \label{sec:inference}
Our inference pipeline is an ensemble of three models: a snapshotter\cite{hermes}, a whitener, and the neural-network itself. Clients send streaming updates of strain data to a snapshotter. The snapshotter sends the latest state to the whitening module. Finally, batches of whitened data are constructed and analyzed by \texttt{Aframe}, producing predictions. The length of the state maintained by the snapshotter is determined by the length of the timeseries used to estimate the PSD, the batch size, and the inference sampling rate. For our analysis, the whitening module uses the first 64 seconds of the snapshotter state to estimate the PSD and build a whitening filter. The remaining data is whitened, and half a second is cropped from both edges to remove the effects of filter settle-in. The whitened data is then unfolded into a batch of overlapping windows. We use a batch size of 128 windows, and, as an inference sampling rate of 4\,Hz was used, each 1.5\,s window overlaps its neighbors by 1.25\,s. This batch of windows is passed to the neural-network for prediction. Lastly, neural-network predictions are aggregated client-side and post-processed via the integration and clustering described above.

\begin{figure}[h]
\centering
\includegraphics[width=\columnwidth]{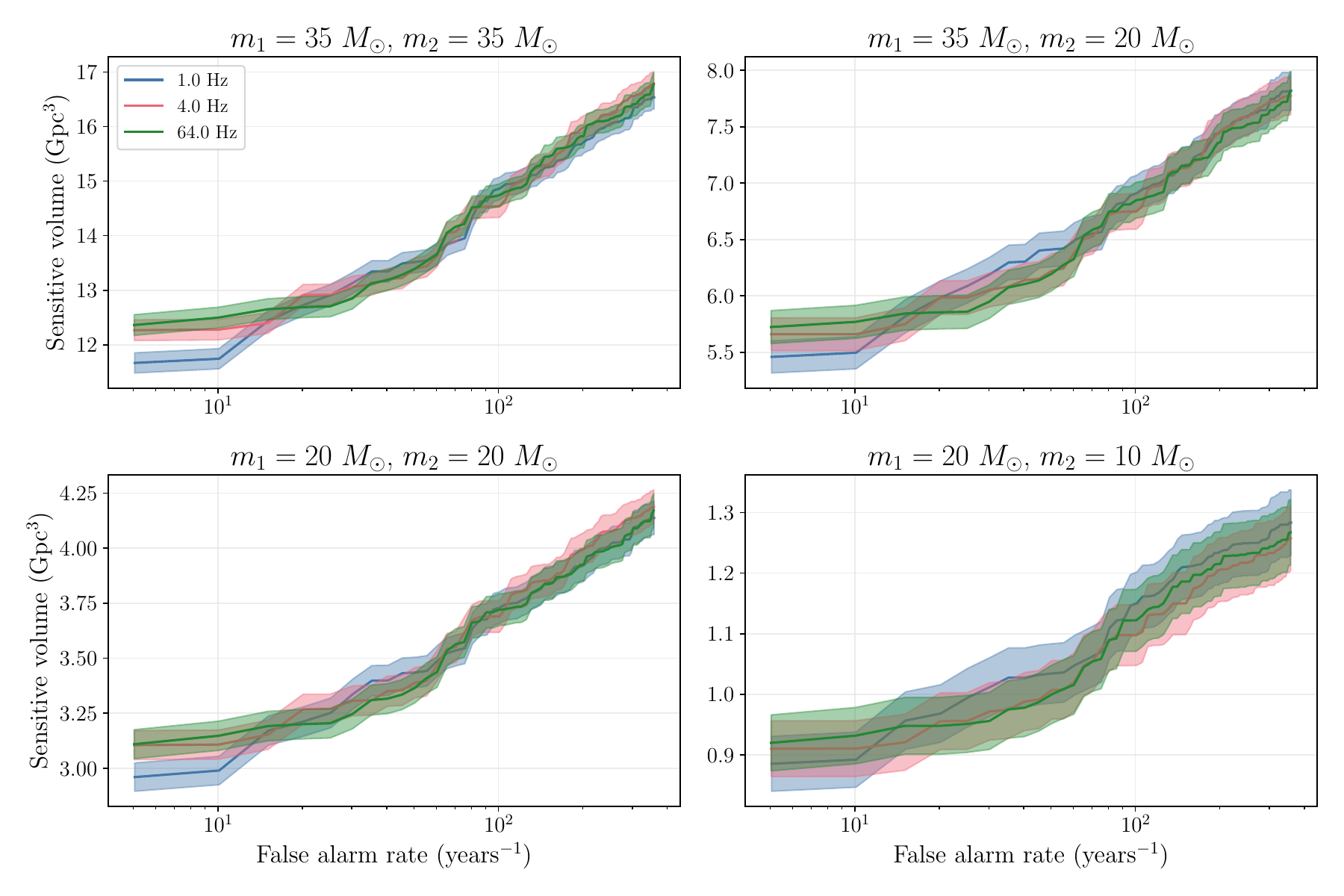}
\caption{ Sensitivity comparisons for the same neural-network run over the same data at different inference rates. For the purposes of clarity, only a subset of the tested rates are shown here. Except for the 1\,Hz inference, all results are within error of each other for all mass combinations and FARs, including for rates not shown in this plot.}
\label{fig:inference-rate-comparison}
\end{figure}

A critical parameter is the inference sampling rate. The inference sampling rate controls the stride between consecutive windows seen by the neural-network. Too small of an inference sampling rate, and astrophysical events may be skipped over. Too large, and computing resources are wasted on redundant inferences. We examined the impact of the inference sampling rate on our sensitivity by repeating trials of our inference procedure at inference sampling rates of 1, 2, 4, 8, 16 and 64\,Hz. For this analysis, we accumulated two months worth of timeslide data for each trial.
Fig.~\ref{fig:inference-rate-comparison} shows a subset of the results of this analysis. Algorithms mostly perform within their statistical error. However, at low FARs the 1\,Hz analysis has a small performance dip in the 35-35 mass bin. Because analyses performed at 4\,Hz require 16 times fewer inference requests than 64\,Hz without sacrificing performance, we use an inference sampling rate of 4\,Hz for the analyses in this paper.

\section{Algorithm Longevity} \label{sec:longevity}

\begin{figure*}[ht]
\centering
\includegraphics[width=\textwidth]{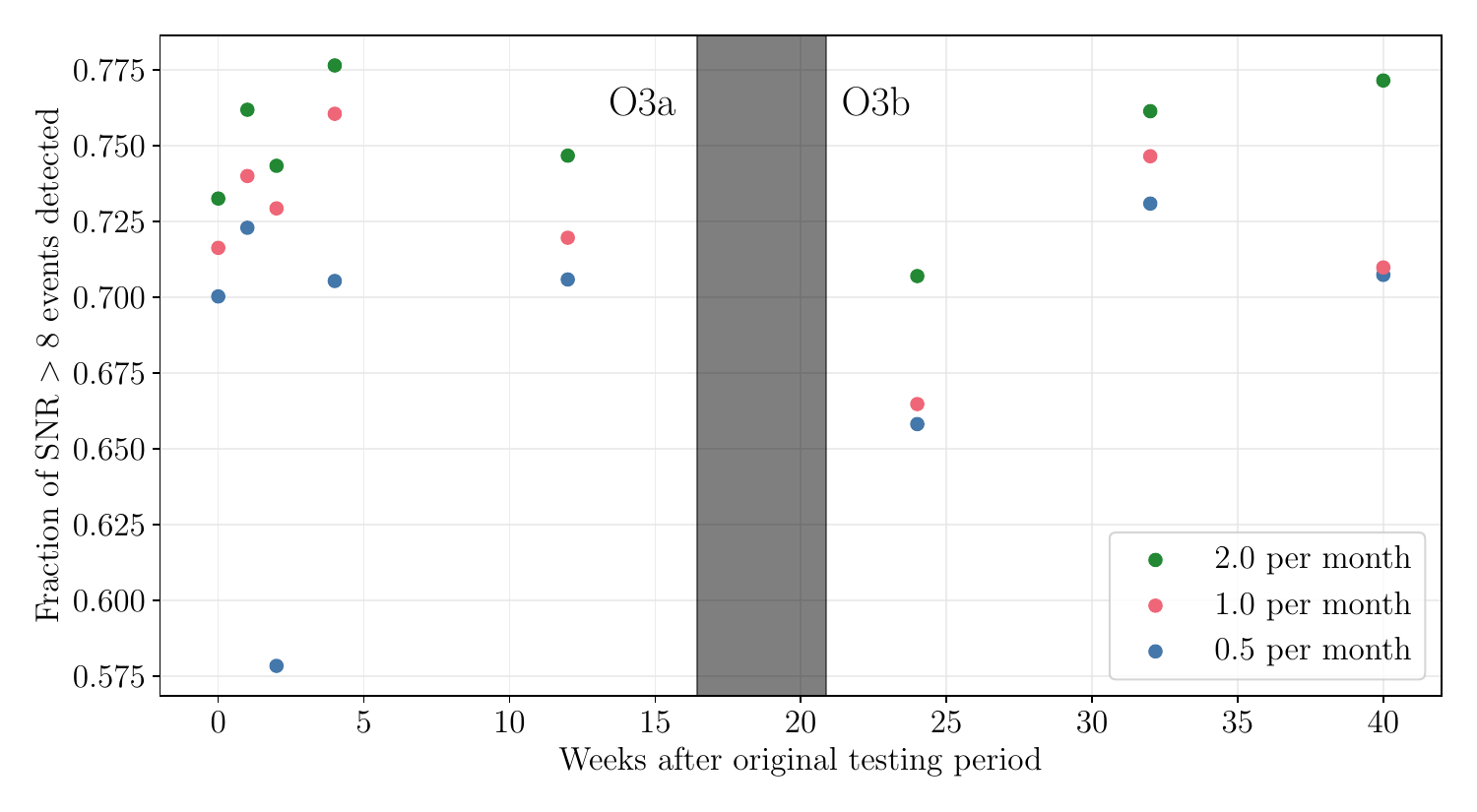}
\caption{The fraction of SNR $>$ 8 events detected at different false alarm rates during various weeks across a period of time during O3, beginning May 9th, 2019 and ending March 21st, 2020. Errors on detection fraction estimates are smaller than the plotted points.}\label{fig:sensitivity_over_time}
\end{figure*}

Noise in gravitational wave interferometers is non-stationary. Therefore, the timescale over which a single trained neural-network will maintain its originally measured performance needs to be evaluated. Determining this timescale helps inform the cadence at which retraining is needed, if at all. 
To test the longevity of our algorithm, we construct several testing datasets at various intervals across O3. For each interval, we analyze the testing dataset with a neural-network trained using the first 10 days of O3 data. This is the same neural-network used to produce the results in Fig.~\ref{fig:sensitivity}. To separate the sensitivity of the neural-network from the sensitivity of the detectors, we do not measure sensitive volume, but instead look at the fraction of events with SNR $>$ 8 that are detected at different FARs. We choose this SNR threshold as it is a typical value often used in GW literature for detectable sources. This metric takes into account the variation in noise level across different time periods, though it does not account for all aspects of detector performance, such as the rate or morphology of glitches. At a FAR of 1 event per 2 months, a threshold comparable to the 1 event per 5 months used for releasing significant public alerts by the LVK\footnote{\url{https://emfollow.docs.ligo.org/userguide/analysis/}}, we see in Fig~\ref{fig:sensitivity_over_time} that the fractional detection rate of the original neural-network does not decay with time. We note that the most significant noise event across all weeks is found during week 2, corresponding to the sharp drop in detection fraction at a FAR of 1 per 2 months. Though there is some fluctuation from week to week, a single neural-network trained on a week's worth of data at the beginning of the observing run maintains sensitivity over the duration of the run. 

\section{Latency and Computational Requirements} \label{sec:compute}
Training the neural-network with a single NVIDIA 16\,GB Tesla V100 GPU takes approximately 43 hours, and once trained, the neural-network can continue to be used for months without retraining; see the discussion of algorithm longevity in Sec.~\ref{sec:training} for details. For inference, we utilize a Triton inference server\cite{NVIDIA_Corporation_Triton_Inference_Server} that is hosted on a NVIDIA DGX server containing eight 16\,GB Tesla V100 GPUs (See Sec.~\ref{sec:inference} for details on inference configuration). Altogether, analyzing the one year of background data and one year of injections used in this analysis to create Fig.~\ref{fig:sensitivity} takes approximately 4 hours, corresponding to a throughput of about 500 seconds of data from a two detector network analyzed per second per GPU. Normalizing by the quantity of GPUs and number of interferometers in the analysis configuration, this corresponds to an order of magnitude improvement in throughput compared with previous work by Huerta et al\cite{huerta2021}, a factor of $\sim2.5$ compared with Chatruvedi et al\cite{Chaturvedi_2022} and a factor of $\sim1.5$ improvement compared with Tian et al\cite{Tian_2024}. This improvement is due to the use of a more efficient neural-network architecture, as well as the IaaS model described above. 

With trained neural-network weights in hand, the requirements for online deployment are much smaller. A single NVIDIA 24\,GB A30 GPU is sufficient for real-time inference at an inference sampling rate of 2048 Hz, which provides sufficient resolution for coalescence time estimation. The total memory required to hold both the neural-network and data is 4.6\,GB. The computational latency of the neural-network is less than 10 milliseconds. In practice, the latency of our algorithm is dominated by pre- and post-processing steps that bring the total latency to approximately 3.1\,s. For a detailed accounting of sources of latency within \texttt{Aframe}, see Table~\ref{table:latency}. The most significant source of latency in an online analysis comes from waiting for data to exist such that data can be cropped from the edges after resampling and whitening. All other computational steps (data reading/writing, data transfer to/from GPU, whitening, event identification, etc.) take less than 0.4 seconds combined. In production, additional latency is incurred uploading events to the GRAvitational-wave Candidate Event
DataBase (GraceDB) \footnote{\url{https://gracedb.ligo.org/}}. This latency is not included in this 3.1\,s estimate. A recent study\cite{chaudhary2023lowlatency} used a real-time mock data challenge replay of O3 data to benchmark pipeline latencies, including GraceDB processing. Analyzing this data stream, we find a median (90\%) event reporting latency of 3.9\,s (4.3\,s), in good agreement with our latency budget. Matched filtering pipelines report a median (90\%) latency of 12.3\,s (41.4\,s).

\begin{table}
\centering
\setlength\tabcolsep{8pt}
\begin{tabular*}{\columnwidth}{ p{5cm} p{3.5cm} }
\hline
\hline
Latency Source & Latency (s)\\
\hline
Coalescence point exiting training kernel padding & 0.25\\
Cropping corruption from whitening filter & 0.50\\
Cropping corruption from resampling to 2048\,Hz & 1.0\\
Integrating network output & 1.0\\
\hline
Reading data and transferring to GPU & $1.03^{+0.06}_{-0.05} \times 10^{-2}$\\
Estimating PSD and whitening & $8.77^{+1.35}_{-0.31} \times 10^{-4}$\\
Performing inference on whitened data & $9.63^{+0.38}_{-0.32} \times 10^{-3}$\\
Integrating and aggregating network output & $3.42^{+0.02}_{-0.01} \times 10^{-1}$\\
Identifying candidate events in integrated output & $1.40^{+0.62}_{-0.43} \times 10^{-4}$\\
\hline

Total & $3.114^{0.006}_{0.001}$\\
\hline
\hline
\end{tabular*}
\caption{Sources of latency for an \texttt{Aframe} online analysis. For the items listed in the upper section this table, the latency does not come from performing the computation, but rather from needing to wait for the data to exist before the action can occur. Items in the lower section are computational steps, and we report the median timing of 9191 trials. The upper and lower error bars represent the 95th and 5th percentile, respectively. All measurements were taken on a dedicated NVIDIA A30 GPU.}
\label{table:latency}
\end{table}

\section{Comparison with Existing Searches}\label{sec:comparison}
To demonstrate \texttt{Aframe}'s readiness for real-time deployment, we compare its sensitivity to search pipelines used in production by the 
LVK. For our pipeline, estimating sensitive volume requires analyzing simulated GW events ``injected'' into strain data, and analyzing background livetime produced by ``timeslides.'' Performing timeslides is a standard way of empirically estimating the background (i.e. the distribution of noise events) for a search pipeline which analyzes a network of detectors. In brief, the strain from one detector is shifted in time by an amount greater than the gravitational wave travel time between the detectors ($\sim$\,10\,ms for the two LIGO detectors). Therefore, any reported triggers could not have been caused by an astrophysical event. In this analysis, the Hanford strain data is held fixed and the Livingston data is shifted in 1\,s increments until the required background livetime is accumulated. Then, a false alarm rate (FAR) can be assigned to injected events by dividing the number of noise events with detection statistic greater than the event of interest, with livetime analyzed. All GW detections reported in the third Gravitational-Wave Transient Catalog (GWTC-3)\cite{Abbott_2023} were excised from the background. 

\begin{figure*}
\centering
\includegraphics[width=\textwidth]{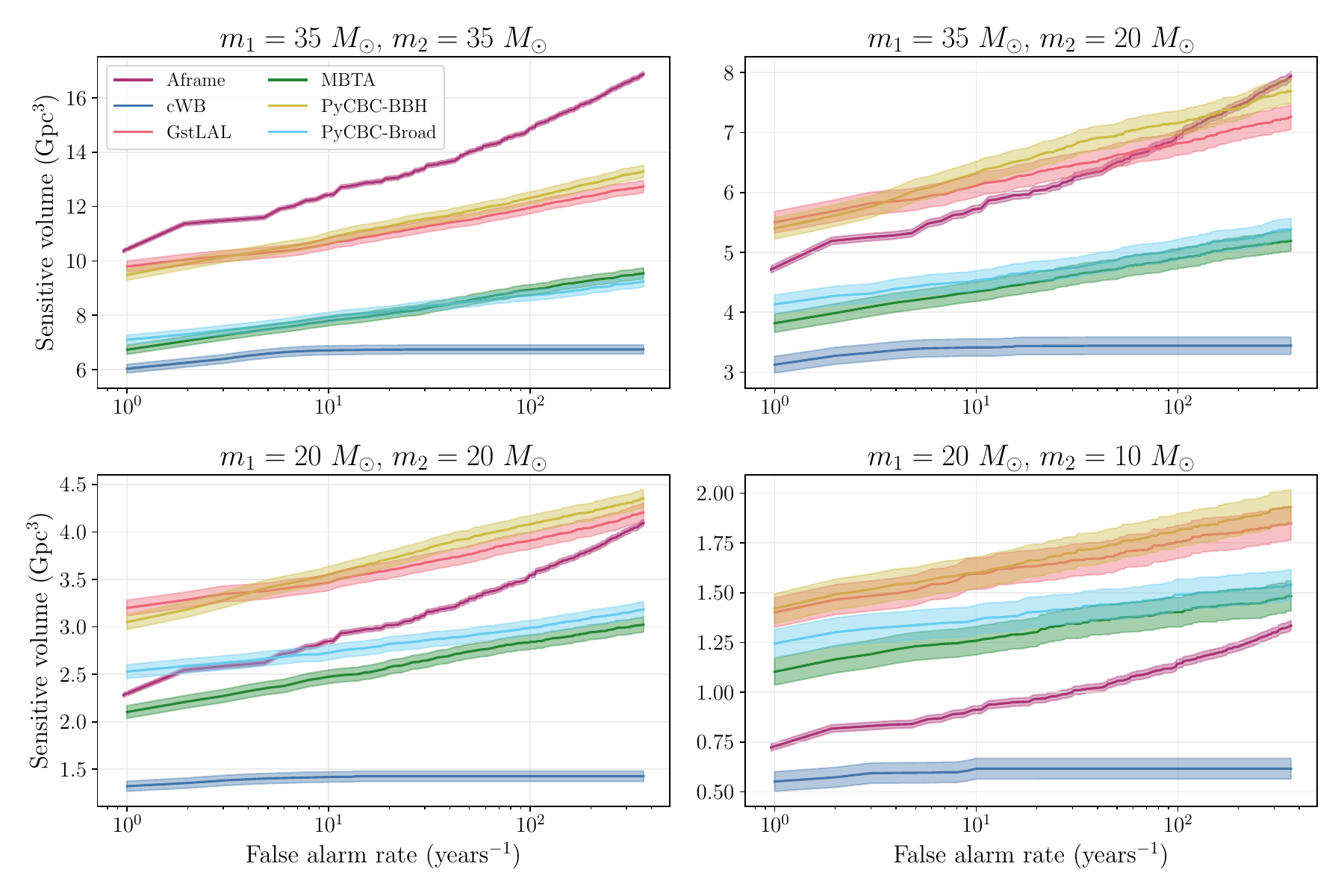}
\caption{Sensitive volume vs FAR for four different mass distributions. Masses are specified in the source frame. Each mass is drawn from a log-normal distribution with a mean of the value given above each plot and a width of 0.1. \texttt{Aframe} demonstrates state-of-the-art sensitivity at higher masses, but loses performance relative to traditional search pipelines at lower masses. The sensitive volume of the other pipelines was calculated using data from a GWTC-3 data release\cite{ligo_scientific_collaboration_and_virgo_2023_7890437}.}\label{fig:sensitivity}
\end{figure*}

A useful metric to measure the sensitivity of search algorithms is the \emph{sensitive volume}. Sensitive volume measures the volume over which some astrophysical population of sources distributed uniformly in co-moving volume is detectable at a given false alarm rate (FAR). Sensitive volume was used to measure the sensitivity of search pipelines in GWTC-3. This provides an astrophysically meaningful benchmark to compare the performance of \texttt{Aframe} to the performance of traditional searches. More details on the sensitive volume calculation can be found in Sec.~\ref{sec:sv-calc}. Fig.~\ref{fig:sensitivity} compares \texttt{Aframe}'s sensitive volume as a function of FAR with the sensitivity of the MBTA\cite{Aubin_2021}, PyCBC\cite{Dal_Canton_2021}, GstLAL\cite{ewing2023performance, Tsukada_2023} and cWB\cite{DRAGO2021100678} searches as reported in GWTC-3\cite{Abbott_2023}. We note that the template banks used by MBTA, GstLAL, and PyCBC-Broad in the GWTC-3 analysis contain waveforms outside of the 5--100 M$_{\odot}$ range searched by \texttt{Aframe}. In principle, these searches could increase their sensitivities in the 5--100 M$_{\odot}$ range by removing these templates. This is evident when comparing the performance of PyCBC-BBH and PyCBC-Broad in Fig.~\ref{fig:sensitivity}. For the future, we encourage production level LVK CBC pipelines to publish BBH-specific sensitivities against which developing ML pipelines can benchmark.

In the 35-35 M$_{\odot}$ mass distribution, \texttt{Aframe} has a larger sensitive volume than the GWTC-3 configurations of all searches, and is comparable in the 35-20 M$_{\odot}$ mass bin, for the FARs considered in this analysis. As source masses decrease further, so does \texttt{Aframe}'s performance relative to existing pipelines. This is in part due to our neural-network architectures inability to model the longer duration features of these low mass signals. While the architecture implements global pooling layers, the convolution layers use a kernel length of 3 samples. Improvements to neural-network architecture design, such as utilizing dilated convolutions that can better model these features will help to improve performance at these mass ranges.

\begin{figure}
\centering
\includegraphics[width=\columnwidth]{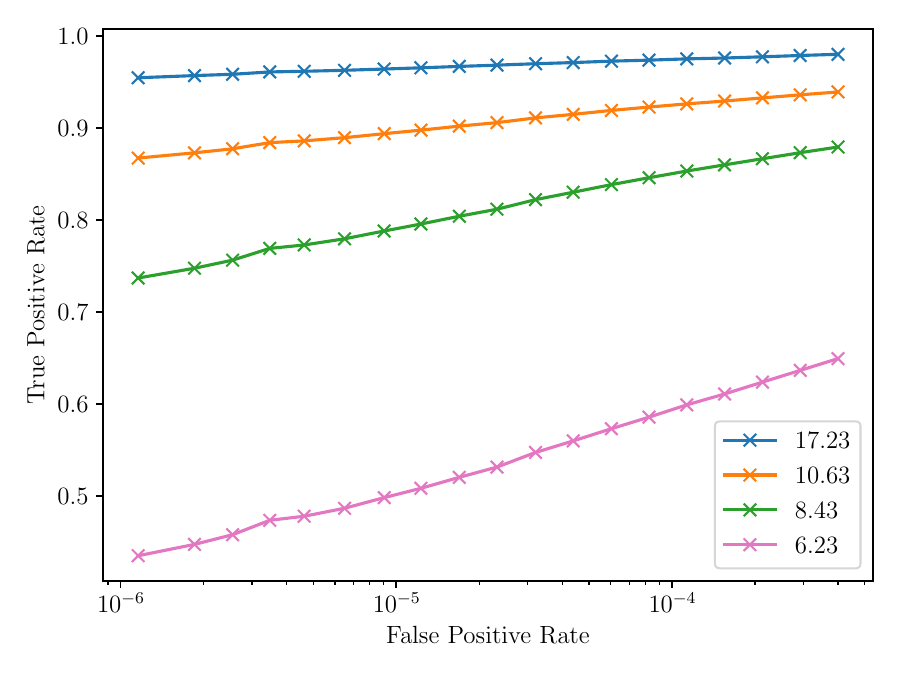}
\caption{ROC curves for waveforms in different SNR bins in our testing dataset, described in Sec.~\ref{sec:data}. Each bin contains waveforms with SNRs within 0.01 of the given value.}\label{fig:roc}
\end{figure}

Previous studies of ML-based gravitational wave detection algorithms tend not to use sensitive volume as a metric, preferring instead to use traditional ML metrics such as ROC curves (an exception is\cite{ml_mdc}, which uses a non-astrophysical prior and a Euclidean volume distribution). This makes direct comparison difficult, as these metrics depend on the parameter distributions of tested events. Still, in Fig.~\ref{fig:roc} we present our own ROC curve to directly compare to metrics published in previous works \cite{huerta2021, Chaturvedi_2022}. We report a higher true positive rate (TPR) for all SNR thresholds at the lowest false positive rate (FPR) of $\sim10^{-6}$ compared to these previous studies. This is particularly true for the lowest SNR threshold of 6.23, where we achieve two orders of magnitude of improvement (a TPR of 0.053, compared to approximately $5\times10^{-4}$). However, we encourage future studies to use sensitive volume to astrophysically motivated distributions as the measure of performance.

\section{Detecting Astrophysical Candidates in GWTC-3.} \label{sec:catalog-events}
The testing period we use contains 9 astrophysical candidate events reported as significant detections in GWTC-3. While we evaluated our algorithm's performance using ``timeslides'' of this data (see Sec.~\ref{sec:comparison}), we also analyzed the unshifted (or ``zero-lag'') data to determine if our algorithm detects these known candidates. The results of this analysis are shown in Table~\ref{table:known-events}. We detect all 9 candidates, with 8 of the 9 candidates detected at a false alarm rate of less than 1 per year, the minimum possible value for this analysis. For the final event, our reported false alarm rate, 14 per year, is of a similar magnitude to the false alarm rate reported by the GWTC-3 pipelines at 2.8 per year. Additionally, during this period, we do not report any non-catalog candidates with a false alarm rate less than 5 per month.

\begin{table*}
\setlength{\tabcolsep}{5pt}
\centering
\scriptsize
\begin{tabular*}{\textwidth}{ p{2.5cm} p{1.3cm} p{1.3cm} p{1.25cm} p{1.5cm} p{1.8cm} p{1.5cm} p{1.8cm} p{1.8cm}}
\hline
\hline
Event & $m_{1} (M_\odot)$ & $m_{2} (M_\odot)$  &Aframe & cWB & GstLAL & MBTA & PyCBC-BBH & PyCBC-Broad
\\
\hline
GW190512\_180714 & $23.2^{+5.6}_{-5.6}$ & $12.5^{+3.5}_{-2.6}$ & $< 0.97$ & $0.88$ & $< 1.0 \times 10^{-5}$ & $0.038$ & $< 1.1 \times 10^{-4}$ &$1.1 \times 10^{-4}$ \\
GW190513\_205428 & $36.0^{+10.6}_{-9.7}$ & $18.3^{+7.4}_{-4.7}$ & $< 0.97$ & -- & $1.3 \times 10^{-5}$ & $0.11$ & $0.044$ & $19$\\
GW190514\_065416 & $40.9^{+17.3}_{-9.3}$ & $28.4^{+10.0}_{-10.1}$  &$14$ & -- & $450$ & -- & $2.8$ & -- \\
GW190517\_055101 & $39.2^{+13.9}_{-9.2}$ & $24.0^{+7.4}_{-7.9}$ &$< 0.97$ & $0.0065$ & $0.0045$ & $0.11$ & $3.5 \times 10^{-4}$ & $0.0095$\\
GW190519\_153544 & $65.1^{+10.8}_{-11.0}$ & $40.8^{+11.5}_{-12.7}$ &$< 0.97$ & $3.1 \times 10^{-4}$ & $< 1.0 \times 10^{-5}$ & $7.0 \times 10^{-5}$ & $< 1.1 \times 10^{-4}$ & $< 1.0 \times 10^{-4}$\\
GW190521 & $98.4^{+33.6}_{-21.7}$ & $57.2^{+27.1}_{-30.1}$ &$< 0.97$ & $2.0 \times 10^{-4}$ & $0.20$ & $0.042$ & $0.0013$ & $0.44$ \\
GW190521\_074359 &$43.4^{+5.8}_{-5.5}$ & $33.4^{+5.2}_{-6.8}$ & $< 0.97$ & $1.0 \times 10^{-4}$ & $< 1.0 \times 10^{-5}$ & $1.0 \times 10^{-5}$ & $< 2.3 \times 10^{-5}$ & $< 1.8 \times 10^{-5}$ \\
GW190527\_092055 &$35.6^{+18.7}_{-8.0}$ & $22.2^{+9.0}_{-8.7}$ & $< 0.97$ & -- & $0.23$ & -- & $19$ & -- \\
GW190602\_175927 &$71.8^{+18.1}_{-14.6}$ & $44.8^{+15.5}_{-19.6}$ & $< 0.97$ & $0.015$ & $< 1.0 \times 10^{-5}$ & $3.0 \times 10^{-4}$ & $0.013$ & $0.29$ \\
\hline
\hline
\end{tabular*}
\caption{Masses in units of $M_{\odot}$, and false alarm rates in units of inverse years from \texttt{Aframe}, cWB, GstLAL, MBTA, and PyCBC-BBH for the known events in our testing set. Masses come from Table VIII of GWTC-2.1\cite{theligoscientificcollaboration2022gwtc21}, and FARs from Table XV of GWTC-3\cite{Abbott_2023}. As our analysis examined only one year of background, our minimum FAR is one per year.}
\label{table:known-events}
\end{table*}

\section{Conclusion}
We have implemented a machine-learning based CBC search pipeline that is capable of low-latency use in a production setting. Through robust data augmentation techniques and extensive work in developing software infrastructure (Sec.~\ref{sec:software}), our algorithm achieves a sensitivity that is superior to or competitive with established search pipelines for higher mass BBHs. Work remains to improve the algorithm's performance on lower mass BBH systems. We leave these investigations to future work.

There are a number of extensions we plan to investigate in future work. As a means of reducing the scope of this study, our algorithm was trained requiring data from both Hanford and Livingston, limiting analysis to when both of these detectors are online. However, accounting for instances where specific detectors are offline is important for maximizing analysis livetime. This could take the form of four-detector, three-detector, pairwise, and single-detector models. Investigating additional detector combinations is left for future work.
Further, low-latency alerts are less important for BBHs than binary neutron star (BNS) and neutron star-black hole (NSBH) mergers, where electromagnetic counterparts are more likely. 
Many algorithms have already been published on this front\cite{Lin_2019, Sch_fer_2020, krastev2020realtimedetectiongravitationalwaves, Aveiro_2022, Qiu_2023}. However, appreciable sensitivity compared with state-of-the-art matched filtering approaches has yet to be demonstrated.
The detection of these mergers with neural-networks is more challenging due to the greater length of time these signals spend in the sensitive band of the detector.

Finally, because \texttt{Aframe} does not provide an SNR timeseries for detections, it cannot be used in conjunction with current low-latency sky localization algorithms like BAYESTAR \cite{Singer_2016}. However, machine-learning based parameter estimation algorithms, like AMPLFI \cite{amplfi}, are being developed as low-latency solutions to sky localization, and can be used to followup \texttt{Aframe} detections.

\section{Data and Software Availability} \label{sec:software}
All code used to produce results in this work is publicly available. 
The \texttt{Aframe} project repository can be found at \url{https://github.com/ML4GW/aframe}. 

In addition, two open source libraries, \texttt{ml4gw}\footnote{\url{https://github.com/ML4GW/ml4gw}} and \texttt{hermes}\footnote{\url{https://github.com/ML4GW/hermes}} were developed to support this work. The \texttt{ml4gw} library contains PyTorch utilities for efficient on-GPU data-loading, whitening, PSD estimation and other data processing techniques common to GW analysis. The \texttt{hermes} library contains utilities for deploying models in the IaaS paradigm via Triton Inference Servers.
\\
\\

\section{Acknowledgements} \label{ack}
W.B and M.W.C acknowledge support from the National Science Foundation (NSF) with grant numbers PHY-2010970 and PHY-2117997. E.M, E.K, A.G and D.C acknowledge support from the NSF under award PHY-1764464 and PHY-2309200 to the LIGO Laboratory and PHY-2117997 to the A3D3 Institute. E.M, E.K, A.G and D.C also acknowledge NSF awards PHY-1931469 and PHY-1934700

The authors express sincere gratitude to Reed Essick for clarifying numerous nuances of the sensitive volume calculation. 
The authors are also very grateful to Thomas Dent for identifying a public dataset which could be used to calculate the performance of traditional search pipelines.
Additionally, the authors thank Vassilis Skliris and Becca Ewing for their useful comments and discussion on earlier drafts of this work.

This research was undertaken with the support of the LIGO computational clusters.
This material is based upon work supported by the NSF's LIGO Laboratory which is a major facility fully funded by the National Science Foundation. This research has made use of data obtained from the Gravitational Wave Open Science Center (www.gw-openscience.org), a service of LIGO Laboratory, the LIGO Scientific Collaboration and the Virgo Collaboration.

\bibliography{references.bib} 
\bibliographystyle{apsrev}

\end{document}